\documentclass[prl,twocolumn,floatfix]{revtex4}

\usepackage{graphicx}

\begin{document}

\title{Self-referenced prism deflection measurement schemes with microradian precision}

\author{Rebecca Olson}
\altaffiliation{Currently at: Department of Physics, University of
Maryland, College Park, Maryland 20742, USA}
\author{Justin Paul}
\author{Scott Bergeson}
\author{Dallin S. Durfee}
\affiliation{Department of Physics and Astronomy, Brigham Young
University, Provo, Utah 84602, USA}

\date{\today}

\begin{abstract}

We have demonstrated several inexpensive methods which can be used
to measure the deflection angles of prisms with microradian
precision.  The methods are self-referenced, using various
reversals to achieve absolute measurements without the need of a
reference prism or any expensive precision components other than
the prisms under test. These techniques are based on laser
interferometry and have been used in our lab to characterize
parallel-plate beamsplitters, penta prisms, right angle prisms,
and corner cube reflectors using only components typically
available in an optics lab. Published in Applied Optics, Vol. 44,
No. 22. \copyright 2005 Optical Society of America.

\end{abstract}



\maketitle

\section{Introduction}

Reflecting prisms are key components in a variety of optical
instruments. They can be used in place of mirrors to alter the
direction of optical beams. Unlike mirrors, however, prisms can be
used in such a way that the angle through which the beam is
deflected does not change when the optic is rotated.  For example,
after a beam reflects off of the three perpendicular surfaces of a
corner cube it will exit the prism travelling in precisely the
opposite direction as the incoming beam.  No careful alignment is
needed to achieve this nearly perfect 180 degree deflection.
Reflecting prisms are useful in situations where it is difficult
to perform the initial alignment or when it is critical to
maintain a particular beam deflection for a long period of time.
One well known example is the use of corner reflectors for lunar
ranging experiments \cite{Dickey94ll}. Our interest in prisms is
to generate an extremely stable array of laser beams for use in an
atom interferometer.

Since the beam deflection is determined by the angles between the
prism surfaces rather than the alignment of the optic, it is
extremely important that the prisms be made correctly. Several
methods are commonly used to measure deflection angles of prisms
\cite{MalacaraOSTprismsA,MalacaraOSTprismsB,MalacaraOSTprismsC}.
One class of techniques utilizes telescopes and autocollimators to
image the separation of two beams at infinity. Our methods are
based on a second class in which the angle between the two beams
is ascertained using optical interference. Both types of
measurements are limited by the size of the beam of light passing
through the optics, in the first case by Rayleigh's criterion, and
in the second by the large fringe spacing resulting from nearly
parallel beams.  As such, both types of measurements have similar
ultimate resolution limits. Techniques based on either type of
measurement typically require a calibrated reference prism or
other expensive optical components.

After purchasing a set of extremely high precision prisms for use
in an atom interferometer, we began to have doubts as to whether
the manufacturer had met our required specifications. Not having
access to an instrument capable of measuring prism deflection
angles to the necessary accuracy, we developed a set of techniques
which allow prism deflection angles to be measured with accuracies
of a few microradians.  Our scheme is self referencing, requiring
no calibrated prism. In addition to the prisms under test we only
needed several standard-quality mirrors, lenses, and attenuators,
an inexpensive alignment laser, a low quality surveillance camera,
and for some measurements a piezoelectric actuator. We
characterized parallel plate beamsplitters (which generate two
precisely parallel beams), penta prisms (which deflect light by 90
degrees), right angle prisms (which fold light by 180 degrees in
the plane of the prism), and corner cubes.

Our methods utilize optical interferometry and bear similarity to
the Jamin interferometer \cite{MalacaraOSTjamin}.  Like several
other schemes, in our methods the deflection angles of prisms are
determined from the spacing between fringes formed by two
interfering beams. Each of our designs produce similar intensities
for the two interfering beams, resulting in high-contrast fringes
for maximum sensitivity. Lenses and mirrors are only used before
the beams are split or after the interference pattern is formed
such that alignment or wavefront errors due to these optics have a
negligible effect on the measurements.

\section{Measuring the Angle Between Two Beams}

When two monochromatic plane waves intersect they form an
interference pattern. Because the spacing between interference
fringes depends on the angle between the two wave vectors, it is
possible to ascertain the angle between the two propagation
directions by analyzing the fringe pattern. Using Fig.\ 1 and
simple trigonometry it is easy to find a relationship between the
fringe spacing $d$ and the angle between the $k$-vectors of the
two plane waves $\Delta \theta$. In the small angle approximation,
for two plane waves with wavelength $\lambda$ projected onto a
screen at near normal incidence, the angle between the two beams
is given by
\begin{equation}
    \Delta \theta = \frac{\lambda}{d} \label{eq:d}.
\end{equation}

\begin{figure}
\begin{center}
\includegraphics{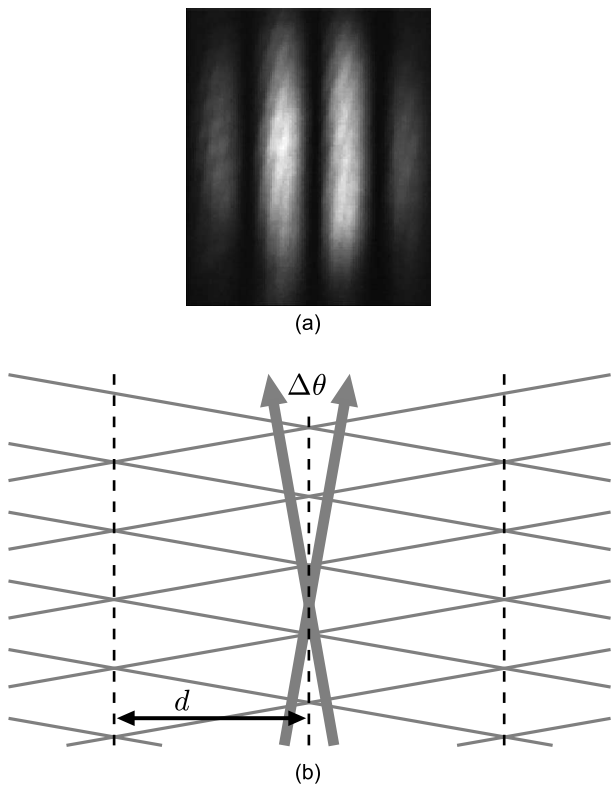}
\end{center}
\caption{Fringes formed by two crossing plane waves.  In (a) the
interference pattern formed when two beams from a HeNe laser
crossed at a small angle is shown. In (b) the gray arrows
represent the propagation or $k$-vectors of the two plane waves,
and the thin gray lines represent the wavefronts of the two
travelling waves. The angle between the propagation vectors of the
two beams is labelled as $\Delta \theta$.  The interference
maxima, where the two waves are always in phase, are denoted with
the dashed lines and the spacing between interference maxima is
labelled as $d$.}
\end{figure}

\subsection{Fitting to a Piece of a Fringe}

Most of the optics we tested have a clear aperture of 2.5 cm. To
prevent clipping we made our measurements using a helium-neon
laser ($\lambda = 632$ nm) collimated to a diameter of about 1 cm,
suggesting that we would only be able to measure fringe spacings
if the fringes were less than 1 cm apart. According to Eq.\
\ref{eq:d} this limit on $d$ results in a minimum measurable
$\Delta \theta$ of 0.13 milliradians. The optics we measured were
specified to have angular tolerances of a few microradians. In
order to make measurements with microradian precision we had to
infer angles from images which contained much less than one
fringe.

One method commonly used in this situation is phase shifting, in
which intensity is measured at several points as the fringe
pattern is scanned across the points by shifting the phase of one
beam \cite{MalacaraOSTpshift}. This method has several advantages
over the spatial fringe-fitting method used in our experiments: it
is less susceptible to wavefront distortion, it reveals the sign
of the angle between the beams (not just the magnitude), and it
can be used for other types of measurements (such as surface
profiling) which cannot easily be done with the method we chose.
But our spatial fringe-fitting method has the advantage that all
of the data is recorded in a single moment, making it more robust
in noisy environments. It also doesn't require the incorporation
of a phase-shifting device, reducing cost and complexity and
eliminating potential errors due to phase shifter beam
deflections, drifts, and hysteresis. Adding a phase shifter would
have greatly complicated our scheme for the measurement of plate
beamsplitters. In our other schemes it could have been implemented
by scanning one prism with a piezoelectric actuator. As discussed
later, we used piezoelectric actuators in some of our schemes for
other purposes, but the piezos lacked sufficient stability for
this purpose (see Fig.\ 7(a) in section \ref{sec:pentaprisms}).

To find the angle between the two beams we simply curve-fit the
intensity pattern on our camera. But to get accurate results when
less than one fringe is visible we have to take into account the
spatial profiles of the beams. To do this we first write down the
expression for the electric field of a laser beam as a function of
the position on the camera $\mathbf{r}$ and time $t$. To simplify
our analysis we assume that the interfering beams have the same
polarization. We also assume that the two beams are well
collimated such that the phase of each beam's electric field is of
the form $\mathbf{k} \cdot \mathbf{r} - \omega t + \phi$ where
$\mathbf{k}$ is the wave vector of the beam, $\omega$ is the
angular frequency of the light field, and $\phi$ is a constant
phase offset. With these assumptions, the electric field of each
beam can be written as
\begin{equation}
    E_n (\mathbf{r},t)  =  f_n(\mathbf{r}) \cos(\mathbf{k}_n \cdot \mathbf{r} - \omega t +
    \phi_n),
\end{equation}
where $f_n (\mathbf{r})$ is the amplitude of the electric field at
position $\mathbf{r}$ and the subscript $n$ is equal to 1 or 2
depending on which of the interfering beams we are describing.

The intensity of the interference pattern of two intersecting
beams is related to the time average of the square of the sum of
the two interfering electric fields. When the time average is
evaluated and the equation is simplified it can be expressed as
\begin{equation}
    I_{12}(\mathbf{r})
    = I_1 + I_2 + 2 \sqrt{I_1 I_2}
    \cos( \mathbf{k}_{\mathrm{rel}} \cdot \mathbf{r} + \Delta \phi),
    \label{eq:simplified}
\end{equation}
where $\Delta \phi = \phi_1 - \phi_2$ and
$\mathbf{k}_{\mathrm{rel}} = \mathbf{k}_1 - \mathbf{k}_2$, and
where $I_1$ and $I_2$ are the intensity patterns which would be
measured on the camera if only one of the two interfering beams
was present.

Without losing generality we can define the plane of the camera's
detector to be the $z=0$ plane (such that $\mathbf{r}$ has no $z$
component). Then we can write the dot product
$\mathbf{k}_{\mathrm{rel}} \cdot \mathbf{r}$ as $k_x x + k_y y$
where $x$ and $y$ are cartesian coordinates describing the
location of pixels on our camera and $k_x$ and $k_y$ are the
spatial frequencies of the interference pattern imaged by the
camera. If both beams strike the camera near to normal incidence,
then the $z$ component of $\mathbf{k}_{\mathrm{rel}}$ will be
nearly zero and $k_{\mathrm{rel}}$ will be approximately equal to
$(k_x^2+k_y^2)^{1/2}$. These definitions result in the following
expression:
\begin{equation}
        \frac{I_{12} - I_1 - I_2 }{2 \sqrt{I_1 I_2}} = \cos \left( k_x x + k_y y + \Delta \phi
        \right).
    \label{eq:cos}
\end{equation}
The left-hand side of this equation can be thought of as a
``normalized'' intensity.

If the two beams are nearly parallel it can be shown that
$k_{\mathrm{rel}} \simeq 2 \pi \Delta \theta / \lambda$. To find
$\Delta \theta$ we simply measure $I_{12}$, $I_1$, and $I_2$ and
numerically fit the left side of Eq.\ \ref{eq:cos} to the right
side to find $k_x$ and $k_y$ treating $\Delta \phi$ as a free
parameter.  We then calculate $k_{\mathrm{rel}}$ and from that
$\Delta \theta$. The three intensity patterns needed to calculate
the left side of Eq.\ \ref{eq:cos} are measured by taking four
images: one of the two interfering beams, one of beam 1 with beam
2 blocked, one of beam 2 with beam 1 blocked, and a ``dark field''
image with both beams blocked.  An example of a set of images is
shown in Figs.\ 2(a)-(d). We then subtract the dark field image
from the other three to generate the three background-free
intensity patterns $I_{12}$, $I_1$, and $I_2$. The separate $I_1$
and $I_2$ terms in Eq.\ \ref{eq:cos} make this measurement
technique work even if the fringes have low contrast due to
mismatched power in the two interfering beams. Lower contrast does
increase digitization noise, which is of special importance when
using a low bit-depth camera. This technique also works if the
interfering beams do not overlap perfectly, although misalignments
can reduce the region of useful data (see Fig.\ 2(e)). Large
overlap misalignments coupled with wavefront curvature in the
beams can also add errors to the measurements.

\begin{figure}
\begin{center}
\includegraphics{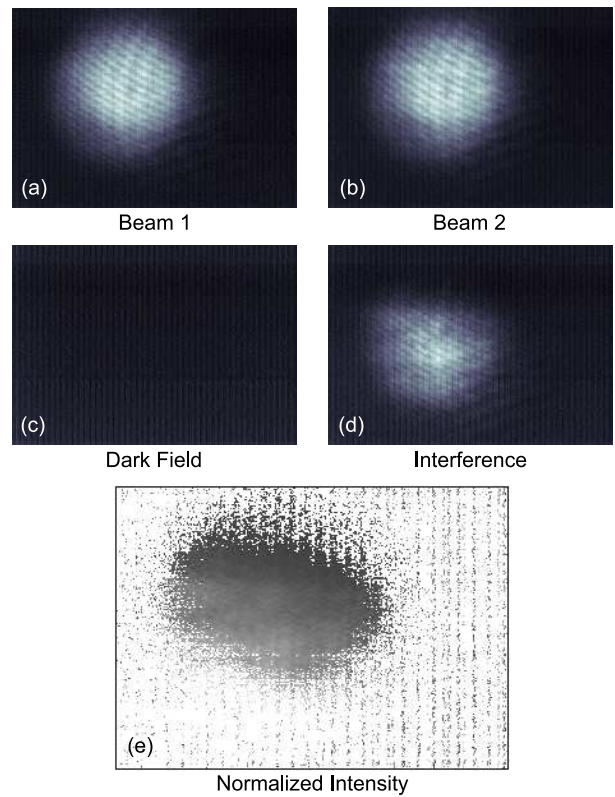}
\end{center}
\caption{Calculating the normalized interference pattern
intensity. Images (a) through (d) are an example of the four
images which are needed to evaluate Eq.\ \ref{eq:cos}. The closely
spaced interference lines visible in these images are low contrast
fringes due to reflections off of the camera window and the
focusing lens.  The high contrast fringes due to the angle between
the two beams are not apparent in the ``interference'' frame
because the spacing between fringes is larger than the size of the
beams.  Plugging the data from these images into the left-hand
side of Eq.\ \ref{eq:cos} results in the image shown in (e). Only
the central part of (e), where both beams are present, contains
meaningful information. The shading scale in (e) runs from -1.35
(pure black) to 0.05 (pure white).}
\end{figure}

Figure 2(e) shows the result of this calculation applied to the
data in Figs.\ 2(a)-(d). Curve fits to find $k_x$ and $k_y$ from
this data are illustrated in Figs.\ 3(a) and (b). Although the
data in these figures is somewhat noisy we can still get accurate,
repeatable results by applying the constraint that the
``normalized interference pattern'' on the left-hand side of Eq.\
\ref{eq:cos} should oscillate with unity amplitude and zero
offset. This is clearly evidenced by the consistency of the
measurements shown in Fig.\ 7 in section \ref{sec:schemes}.

\begin{figure}
\begin{center}
\includegraphics{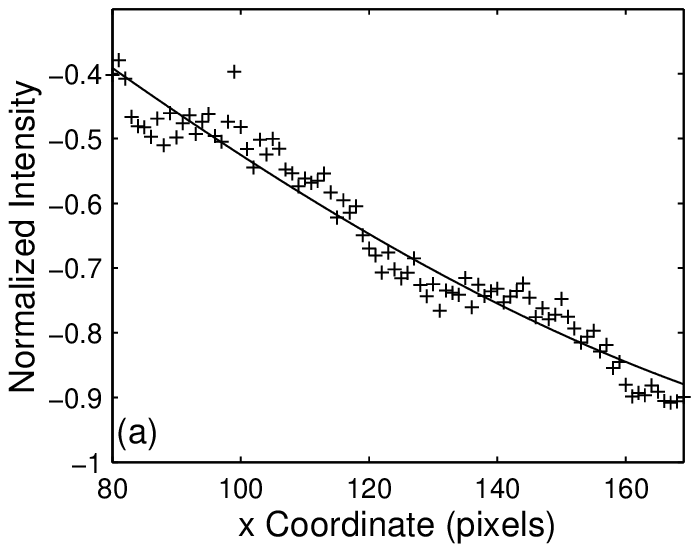}
\includegraphics{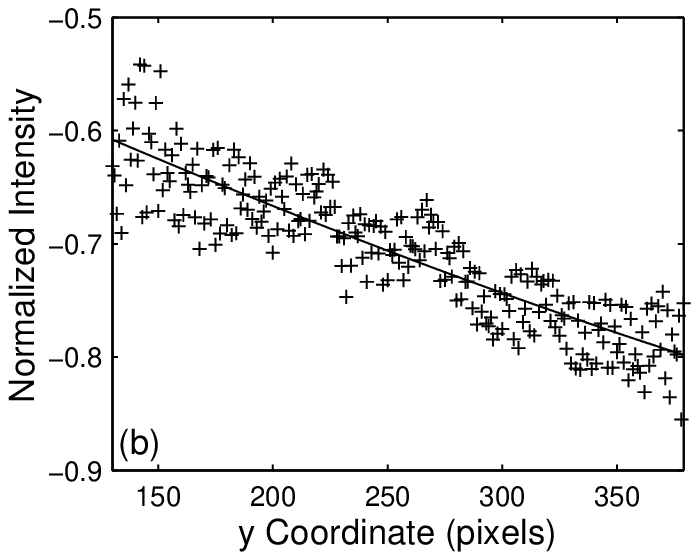}
\end{center}
\caption{Curve fits to find the angle between two beams. Strips
through the center of the data from Fig.\ 2(e) are shown, along
with least-squared fits to the functions $\cos( k_x x + \phi_x )$
and $\cos(k_y y + \phi_y )$. The deviation of the data from the
fits is largely due to camera window reflections.  These higher
spatial frequency low-contrast fringes average away to a large
extent in the curve fit.}
\end{figure}

\subsection{Experimental Subtleties}

When using this curve-fitting approach to measure deflection
angles of prisms we often made small adjustments to the prism or
beamsplitter alignment in order to shift the relative phase of the
two interfering beams such that images were not centered on a
light or dark fringe. Only small adjustments which did not affect
the overlap of the interfering beams were needed. Capturing data
between a light and a dark fringe results in a more precise fit to
the data. Fitting data near an extremum of the cosine requires
precise measurement of the curvature of the data. Near a zero
crossing, however, simply extracting the slope of the data is
enough to get a good measurement of $k$.

In our treatment we have assumed a well-collimated laser beam and
have ignored effects of wavefront curvature.  To ensure good beam
collimation we constructed a simple Michelson-Morley
interferometer with mismatched arms, one arm being about 2
centimeters long and the other over one meter long.  The
interferometer was aligned to create a circular interference
pattern.  We then adjusted the the lenses used to telescope up the
size of the laser beam until no interference rings were visible.
When measuring prism deflection angles we made sure that the two
optical paths were the same length on a millimeter scale and that
the two interfering beams hit the camera at nearly the same place.
This made any residual wavefront curvature common to both field
components such that it did not effect on our results.

The detector on the camera used in these experiments was smaller
than the laser beam diameter.  Since catching only part of the
interference pattern limits sensitivity to small relative beam
angles, we used a lens to demagnify the pattern.  To account for
the demagnification and to find the correct ``effective size'' of
the camera pixels we placed a ruler in front of the lens.  The
ruler's position was adjusted until it came into clear focus on
the camera.  We then took pictures of the ruler to determine the
magnification due to the lens.  We verified that this had been
done correctly by using the lens' focal length and the distance to
the camera to calculate the position at which we would expect the
ruler to come into focus and the expected magnification.

When we evaluated the left-hand side of Eq.\ \ref{eq:cos}, we had
to be careful to utilize only the parts of the images where
sufficient laser light was present in both beams to avoid large
errors due to division by small numbers (see Fig.\ 2(e)). We
designed our software to prompt the user to select a region of
interest to avoid regions of low intensity. The left-hand side of
equation Eq.\ \ref{eq:cos} is then computed in this region. The
software then fits a horizontal row of data in the middle of the
selected region to the function $ \cos( k_x x + \phi_x )$, and
fits a vertical column of data in the middle of the region to the
function $ \cos( k_y y + \phi_y )$. From these two one-dimensional
fits it calculates $k_{\mathrm{rel}}$ and determines the angle
between the beams.

\section{Measuring Prism Beam Deflections}

\label{sec:schemes}

The following paragraphs discuss several methods which we used to
characterize the properties of parallel-plate beamsplitters, penta
prisms, right angle prisms, and corner cubes.  We tested uncoated
optics. Light intensity was lost due to imperfect transmission
each time a beam entered or exited a prism. Much larger losses
occurred due to missing reflective coatings on the beamsplitters
and the penta prisms (right angle prisms and corner cubes do not
require reflective coatings due to total internal reflection). But
even with these losses we could still saturate the camera.
Balancing the intensities of the two interfering beams was
necessary to achieve high contrast fringes to get the most
accuracy with the fixed bit-depth of our camera. Our methods have
symmetric losses in each beam, resulting in well matched beam
intensities.

\subsection{Absolute Beamsplitter Characterization}

The beamsplitters we measured were uncoated plates of BK7 glass
with parallel surfaces.  As shown in Fig.\ 4, when a laser beam
passes through an uncoated piece of glass, surface reflections
result in multiple beams exiting the glass. We are concerned only
with the beam which passes through without reflecting and the
nearly parallel beam resulting from one reflection from each
surface (labelled 1 and 2 in the figure).  If the two beamsplitter
surfaces are exactly parallel, these two beams will emerge exactly
parallel. Otherwise there will be an angle $\theta$ between the
two exiting beams (see Fig.\ 4).  By measuring $\theta$, the prism
wedge angle $\psi$ can be inferred.

\begin{figure}
\begin{center}
\includegraphics{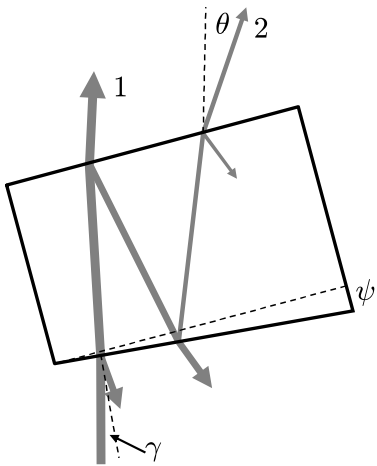}
\end{center}
\caption{Generation of two nearly parallel beams with a plate
beamsplitter.  The gray lines represent laser light. Light enters
the beamsplitter in the lower left-hand corner. At each interface
the beam is split into a reflected and a transmitted beam.  For
most of our studies we are only interested in the two beams
exiting the beamsplitter which are labelled 1 and 2 in the above
figure.  The angle between the incoming beam and the normal of the
first surface is labelled as $\gamma$, the angle between beams 1
and 2 is labelled as $\theta$, and the wedge angle of the glass
plate is labelled $\psi$.}
\end{figure}

The relationship between $\theta$ and $\psi$ can be found using
Snell's law and the law of reflection.  If beam 1 in Fig.\ 4
defines the $z$ axis and the $x$ axis is defined such that the
angle $\gamma$ is in the $x$-$z$ plane, in the limit of small
wedge angles the $x$ component of $\psi$ is related to the the $x$
component of $\theta$ by
\begin{equation}
    \psi_x = \frac{\theta_x}{2} \sqrt{
    \frac{1-\sin^2(\gamma)}{n^2-\sin^2(\gamma)}}
\end{equation}
where $n$ is the index of refraction of the beamsplitter. The $y$
component of $\psi$ is given by the same relationship, with
$\psi_y$ and $\theta_y$ replacing $\psi_x$ and $\theta_x$.

To measure the wedge angle of parallel plate beamsplitters we used
the configuration shown in Fig.\ 5.  In this configuration two
beamsplitters form a Mach-Zehnder interferometer. Since each of
the two beams undergoes two reflections the two interfering beams
have similar intensities, resulting in high contrast interference
fringes.  To get good overlap between the interfering beams and to
make the Fresnel coefficients the same in both beamsplitters, the
two beamsplitters were placed at similar angles relative to the
incoming beam.

\begin{figure}
\begin{center}
\includegraphics{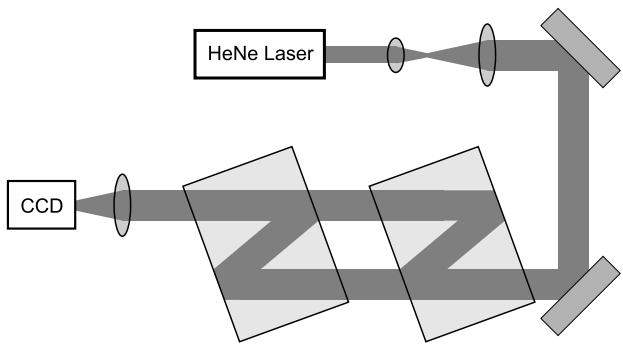}
\end{center}
\caption{Optical setup to measure the wedge angles of
parallel-plate beamsplitters.}
\end{figure}

In this arrangement the interference pattern does not reveal the
wedge angle of a single beamsplitter, but gives a combination of
the wedge angles of both beamsplitters. To find the wedge angle of
a {\em single} beamsplitter we make four measurements using
different combinations of three beamsplitters and use the fact
that flipping a beamsplitter over effectively reverses the sign of
its wedge angle. The first and second measurements use
beamsplitters ``A'' and ``B'' with beamsplitter ``B'' turned over
between measurements. The third and fourth measurements use
beamsplitters ``A'' and ``C'' with beamsplitter ``C'' turned over
between them. In each of the four configurations we measure the
$k_x$ and $k_y$ of the interference pattern to extract the
magnitude of the $x$ and $y$ components of the angle between the
outgoing interfering beams using the methods discussed previously.

If $\theta_{\mathrm{A}x}$, $\theta_{\mathrm{B}x}$, and
$\theta_{\mathrm{C}x}$ represent the $x$ components of the
relative deflection errors of beamsplitters ``A,'' ``B,'' and
``C,'' and the magnitudes of the $x$ components of the angle
between the interfering beams in the four measurements are
represented by $M_{1x}$, $M_{2x}$, $M_{3x}$, and $M_{4x}$, the
four measurements yield the following results:
\begin{eqnarray}
     M_{1x} &=& \theta_{\mathrm{A}x} - \theta_{\mathrm{B}x},s \\
    \pm M_{2x} &=& \theta_{\mathrm{A}x} + \theta_{\mathrm{B}x}, \\
    \pm M_{3x} &=& \theta_{\mathrm{A}x} - \theta_{\mathrm{C}x}, \\
    \pm M_{4x} &=& \theta_{\mathrm{A}x} + \theta_{\mathrm{C}x}.
\end{eqnarray}
A similar set of equations can be written down for the $y$
components.  Fitting our data using Eq.\ \ref{eq:cos} does not
reveal the sign of the angle between the two interfering beams.
But we can assume a convention in which the angle between the two
interfering beams is defined to be positive for our first
measurement.  For the following measurements we must stick to the
same convention. The $\pm$ sign in the lower three relations
therefore results from the uncertainty in the sign of the angle
between the interfering beams when they are measured
interferometrically.

The equations can be solved for the $x$ component of the relative
deflection angle of each beamsplitter as a function of the four
measured angles. But without knowledge of the sign of the angle
between the interfering beams, these expressions cannot be
evaluated. Fortunately, the above system of four equations yields
two independent expressions for $\theta_{\mathrm{A}x}$, one in
terms of $M_{1x}$ and $M_{2x}$, and the other in terms of $M_{3x}$
and $M_{4x}$. In most cases the requirement that
$\theta_{\mathrm{A}x}$ be the same as determined by both equations
unambiguously determines the sign of each measurement term.  Once
the signs are determined, the wedge angle for each of the three
beamsplitters can be determined. Using this technique we
characterized several high-precision beamsplitters, measuring
wedge angles from 1 to 6 $\mu$rad.

\subsection{Relative Penta Prism Characterization}
\label{sec:pentaprisms}

Our application does not place tight requirements on the absolute
angular deflection produced by our penta prisms. It does, however,
require that pairs of penta prisms be precisely matched. As such
we measured the relative deflection of each matched pair rather
than the absolute deflection of individual prisms.  We did this
using the optical configuration shown in Fig.\ 6. In this
configuration one of the plate beamsplitters, characterized using
the methods described above, was used to generate two parallel
beams.  These beams were then folded at right angles using a pair
of penta prisms.  The two beams were then recombined using a
second plate beamsplitter.

\begin{figure}
\begin{center}
\includegraphics{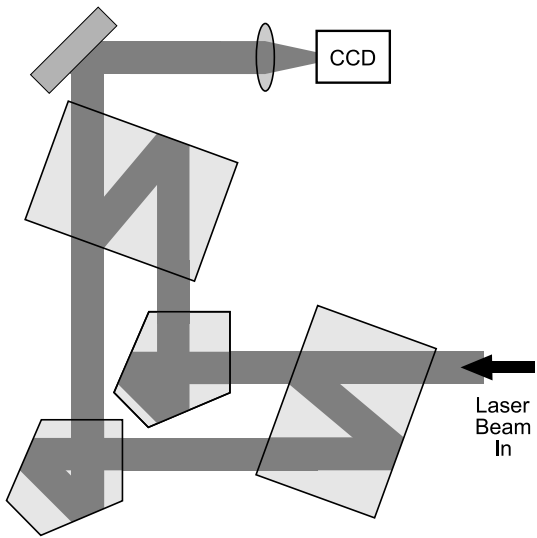}
\end{center}
\caption{Optical setup to measure relative deflection angles of
penta prisms.}
\end{figure}

In this layout the two beam paths are symmetric, allowing us to
make the two path lengths nearly the same and making for equal
intensity losses in each beam as they reflect off of our uncoated
prisms. We used the same angle of incidence for both beamsplitters
to make the Fresnel coefficients equal.  To get the two
interfering beams to overlap we adjusted the separation of the
penta prisms to make the spacing between the two beams entering
the second beamsplitter equal to the spacing of the two beams
exiting the first beamsplitter.

Penta prisms ensure deflection of a beam by a precise angle in the
plane of the prism.  If, however, one prism is tilted out of the
plane defined by the other prism, the two interfering beams would
be at an angle to one another determined not by the accuracy of
the prisms but by their relative alignment. For small
misalignments we can think of the light deflection by the second
prism as a fixed deflection in the plane defined by the first
prism plus an out-of-plane deflection due to misalignment. As
such, the magnitude of the wave vector describing the sinusoidal
interference pattern measured at the output would equal the
quadrature sum of two orthogonal components: a component due to
errors in the manufacture of the prisms and a component due to the
relative alignment of the prisms, as shown in Eq.\
\ref{eq:hyperbola} below.
\begin{equation}
    k_{\mathrm{rel}} = \sqrt{ k_{\mathrm{p}}^2 + k_{\mathrm{a}}^2}
    \label{eq:hyperbola}
\end{equation}
Here $k_{\mathrm{p}}$ represents the component due to error in the
prism, and $k_{\mathrm{a}}$ represents the component due to
alignment error.

Because $k_{\mathrm{rel}}$ is at a minimum when there is no
alignment error (i.e., when $k_{\mathrm{a}} = 0$), it is possible
to measure $k_{\mathrm{p}}$ by making measurements while adjusting
the out-of-plane alignment of one prism.  Rather than searching
for a minimum value, we took several measurements at different
alignments and fit our measurements to the form of equation Eq.\
\ref{eq:hyperbola} to extract an accurate value for
$k_{\mathrm{p}}$. To do this we mounted one of our prisms on a
piezoelectric (PZT) mount which enabled fine alignment
adjustments. We would manually adjust the alignment such that the
minimum of $k_{\mathrm{rel}}$ occurred near the middle of the
range of our PZT actuator.  We then took images as we scanned the
PZT.

Because our fringe analysis method utilizes data taken at a single
moment in time, we were able to make precise measurements of
$k_{\mathrm{p}}$ even though our PZT actuator was unstable.
Assuming that $k_{\mathrm{a}}$ will be proportional to the voltage
$V$ applied to the piezoelectric element, we can take the measured
$k_{\mathrm{rel}}$ as a function of $V$ and perform a curve fit to
find $k_{\mathrm{p}}$. This curve fit requires two free parameters
(in addition to $k_{\mathrm{p}}$): the voltage at which
$k_{\mathrm{a}}=0$ and the constant of proportionality between $V$
and $k_{\mathrm{a}}$.  As shown in Fig.\ 7(a), however, due to
nonlinearity and drift in our piezoelectric mount the data does
not fit the hyperbolic form of Eq.\ \ref{eq:hyperbola} well. But
since the $k_{\mathrm{p}}$ component was approximately in the
horizontal plane of our camera and $k_{\mathrm{a}}$ was in the
vertical, we could perform much better fits when we plotted the
total $k_{\mathrm{rel}}$ vs. $k_y$, the vertical component of
$k_{\mathrm{rel}}$ extracted by our image analysis software. These
fits had no free parameters. Typical curve fits are shown in
figures 7(b) and (c).

\begin{figure}
\begin{center}
\includegraphics{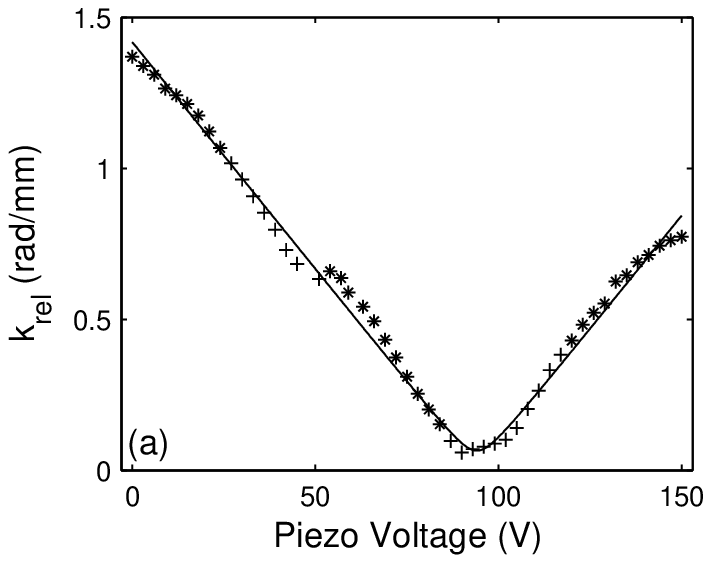}\\
\includegraphics{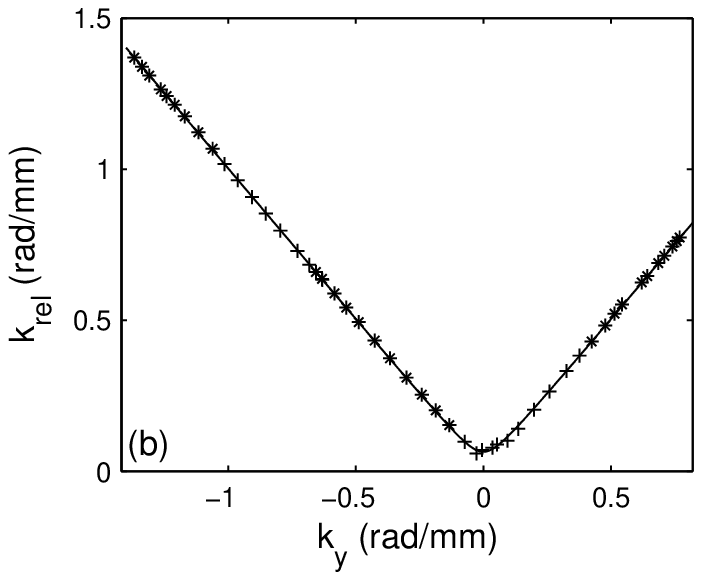}\\
\includegraphics{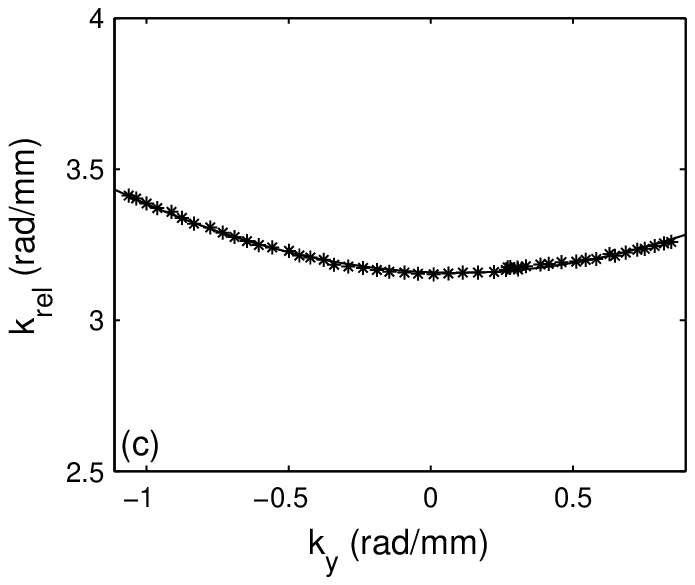}
\end{center}
\caption{Finding the relative deflection error of two penta
prisms. The magnitude of the wave vector describing the
interference pattern at different prism alignments is plotted vs.
the PZT voltage (a) and vs. the $y$ component of the wave vector
(b).  The crosses and the asterisks represent the actual data
extracted from the interference patterns.  The asterisks represent
the data points which should be the most accurate since the image
happened to fall between a light and a dark fringe. The crosses
represent data points for which the image contained a light or
dark extremum. The lines represent equally weighted least-squares
fits of the entire data set to Eq.\ \ref{eq:hyperbola}.  Data from
a different set of prisms which did {\em not} meet our
specifications is shown in (c).}
\end{figure}

The fit in Fig.\ 7(b) yields a $k_{\mathrm{p}}$ of 64.4 rad/m
corresponding to a relative deflection angle of 6.5 $\mu$rad for
the two prisms with an RMS fit error corresponding to 0.38
$\mu$rad. Scanning the PZT had the side effect of moving the
location of bright and dark fringes such that some images
contained an extremum. But comparing the data points in 7(b) which
contained an extremum to those which didn't, it is clear that this
did not significantly reduce the accuracy of the fits. A fit using
just the data for which the image did not contain an interference
minimum or maximum gives a relative deflection of 7.5 $\mu$rad.
Although most of the information in the plots is contained in the
lowest points where the hyperbola is dominated by
$k_{\mathrm{p}}$, simply fitting to the two points at the extremes
of the scan gives a reasonable relative deflection of 6.8
$\mu$rad, implying that only a small number of images are needed
to get accurate results. Similar results were seen for our other
prism pairs suggesting a repeatability of this method at the
$\mu$rad level. Due to the known deflection error of the
beamsplitters used in these measurements, the absolute accuracy of
our measurements was limited to about 2 $\mu$rad.

The consistency of the data in Fig.\ 7 gives a good idea of the
overall accuracy of our fringe measurement technique. One sign of
self-consistency is the fact that the asymptotes of the hyperbola
in Fig.\ 7(b) cross at a value of $k_{\mathrm{rel}}$ which is very
close to zero.  In all of our measurements of precision prism
pairs we measured offsets corresponding to angle measurement
errors ranging from nearly zero to 1.02 $\mu$rad. Another
indication of the accuracy of our fringe analysis is the low RMS
error of the curve fits to Eq.\ \ref{eq:hyperbola}.  These ran
from 0.40 to 1.24 $\mu$rad.

\subsection{Right Angle Prism and Corner Cube Characterization}

We characterized the relative deflection of pairs of right angle
prisms using a scheme similar to the one we used for penta prisms.
Because these prisms deflect light back towards the beamsplitter,
an optical layout analogous with the one we used to measure penta
prisms cannot be used --- a beam reflected off of one prism would
be occluded by the second prism. One approach would be to use a
design in which the beams were deflected vertically back to a
second beamsplitter placed above the first beamsplitter.  To avoid
the complications of multi-tiered optics, we instead used the
layout shown in Fig.\ 8.  In this design a single beamsplitter is
used to split and recombine the two beams. Unlike the schemes
described earlier in this paper, the intensities of the two
interfering beams are not precisely balanced in this setup; while
both paths involve one beamsplitter reflection, the path through
the upper prism undergoes two more transmissions through
beamsplitter surfaces than the path through the lower prism.  Due
to the low reflectivity of the uncoated beamsplitters we still
achieved nearly 100\% fringe contrast.  This same set-up could
also be used to characterize corner cubes.

\begin{figure}
\begin{center}
\includegraphics{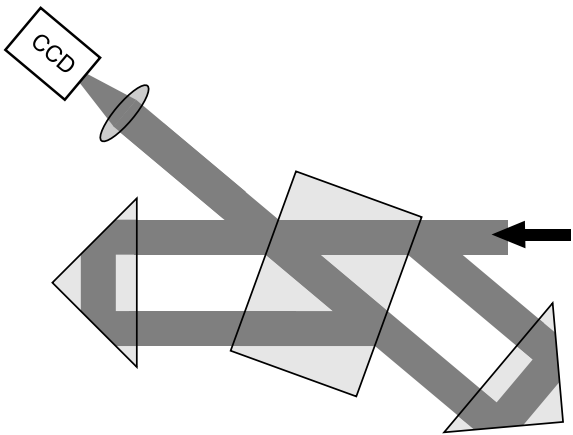}
\end{center}
\caption{Optical setup for measuring the relative deflection of
two right angle prisms or corner cubes.}
\end{figure}

In addition to the two beams we are interested in, a third beam
travelling through the upper prism in the opposite direction can
have an effect on the interference pattern.  This beam undergoes
two additional beamsplitter reflections and is therefore much less
intense.  When measuring right angle prisms, the prisms can be
tilted vertically to walk this stray beam out of the interference
pattern.  With the vertical alignment walked off, the distance
between the beamsplitter and the prisms will have to be adjusted
to achieve good overlap of the interfering beams.  As with the
penta prism measurements, to measure the difference in the
intrinsic deflection angles of two right angle prisms we scanned
the vertical angle of one of the prisms and then fit the measured
relative beam angles to Eq.\ \ref{eq:hyperbola}.  Using this
method we measured the relative deflection angle of pairs of
high-quality right angle prisms.  The repeatability of these
measurements was similar to what we achieved with our penta prism
measurements.

To measure the absolute deflection angle of a single right angle
prism or corner cube we used the scheme illustrated in Fig.\ 9.
Unlike the other schemes presented in this paper this scheme
requires that the beamsplitter angles be chosen carefully. Simpler
designs using one or two beamsplitters had problems with stray
reflections which resulted in interference of more than two paths
and unequal intensities of interfering beams. The three
beamsplitter design allows us to control stray reflections but
requires a different angle of incidence at each beamsplitter. This
results in different Fresnel reflection coefficients at each
beamsplitter. Also, like our scheme for relative measurements of
right angle prisms, in this setup one of the beams undergoes two
more transmissions through a beamsplitter surface than the other
beam.  By carefully choosing the beamsplitter angles one can make
the two pathways overlap and be equal in intensity at the camera.
This is easily done with a knowledge of the beamsplitter thickness
and index of refraction.

\begin{figure}
\begin{center}
\includegraphics[width=7 cm]{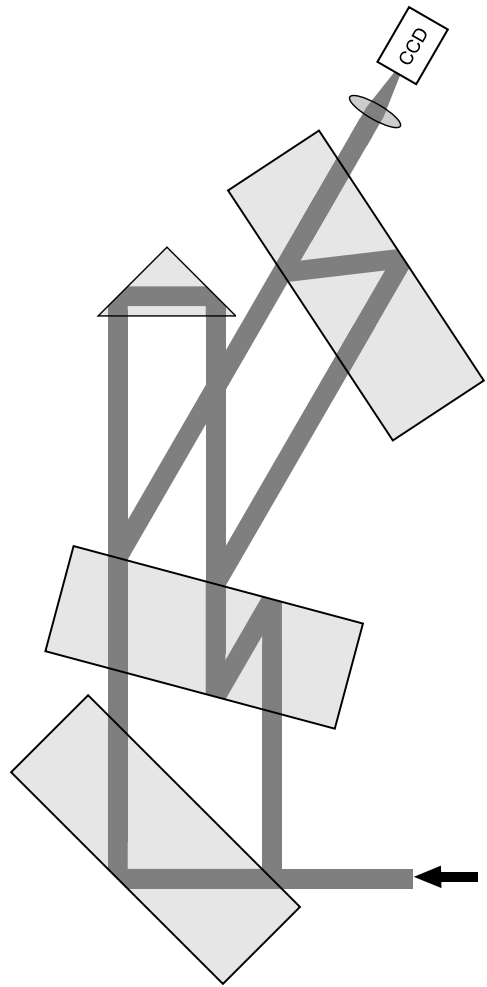}
\end{center}
\caption{Optical setup for absolute measurement of right angle
prism and corner cube deflection angles.}
\end{figure}

We used this method to measure the absolute deflection angles of
several high-quality right angle prisms as well as a low-quality
right angle prism and a high-quality corner cube. The high-quality
prisms and the corner cube deflection angles were typically found
to deviate from 180 degrees by a few microradians. The deflection
angle of the cheap right angle prism was found to be much less
accurate. Once again we found repeatability at the $\mu$rad level.

\section{Components Used}

Our measurements used only the prisms under test and parts
available in our lab. The laser was an inexpensive $\simeq 1$ mW
helium-neon alignment laser [JDS Uniphase Model 1507P]. Because
the prisms and beamsplitters did not have reflective coatings,
only about 0.01 to 1 percent of the laser light reached the camera
depending on the type of prisms being measured. Even so we still
needed significant attenuation to avoid saturating the camera. The
laser had a good spatial mode and a coherence length long enough
to produce good interference fringes on the asymmetric
Michelson-Morley interferometer mentioned previously. A laser with
poorer spatial and temporal qualities could also have been used.
The required spatial mode can easily be achieved by spatial
filtering, especially considering the low power needed. If the two
optical paths are made equal within about 1 mm when measuring the
prisms, a short-term linewidth of tens of GHz would be sufficient
to produce high-contrast fringes. Although a short coherence
length would not allow collimation to be tested using an
asymmetric interferometer, there are many other ways to ensure
good collimation.

The camera was a \$156 closed circuit surveillance camera
connected to a computer frame grabber card.  The low quality
camera resulted in three significant difficulties. First was the
camera's nonlinear response. Our camera was not designed for
scientific work and its response function was not well calibrated.
As a result, in our first measurements the ``cosine'' function in
Eq.\ \ref{eq:cos} did not oscillate between -1 and 1.  We
attempted to characterize the camera's response (surveillance
cameras usually have a response in which the value of each pixel
is proportional to the intensity of light on the pixel raised to
some power $\gamma$). But we found that, even with a fixed iris
setting, at high intensities the signal reported on one pixel
depended on the intensity present on other parts of the chip!  But
for sufficiently low intensities the camera response was fairly
linear.  So our solution was to reduce light intensities by adding
attenuators in front of the camera until the highest value
reported at any pixel was 80 counts (out of a maximum of 255
counts for the 8-bit camera).

The other two problems with the camera were related to its low
signal-to-noise ratio and to an uncoated window on the front of
the camera. The signal-to-noise problem was overcome by averaging
50 frames to produce each image. This took less than 2 seconds on
our 30 frames-per-second video camera.  The uncoated window
affixed to the camera produced low-contrast interference fringes
in our data (see Figs.\ 2 and 3). We were unable to remove this
window. But by tilting the camera we were able to make the spatial
frequency of these fringes high enough that they did not confuse
the fitting routines when fitting the much broader fringes due to
the relative angle of the two interfering beams. Note that Eq.\
\ref{eq:cos} was derived under the assumption that the beams
strike the camera near to normal incidence.  The equation is still
approximately correct when we tilt the camera, especially when the
camera is tilted around an axis which is nearly perpendicular to
the fringes. Tilting our camera, therefore, did not change the way
that we analyzed our images and the residual error due to the
camera tilt was negligible.

The lenses and mirrors we used were standard research-quality
optics which were already available in the lab. Two lenses were
used to telescope and collimate the laser beam before entering the
interferometer. These had to be of reasonable quality to prevent
significant wavefront distortion of the laser. Any distortion due
to these lenses is common to both of the interfering beams and
should have a reduced impact on the measured fringes.  A third
lens was used to demagnify the interference pattern to fit onto
the camera. This lens simply images the interference pattern.
Small wavefront errors at this lens do not have an effect on the
measurement, and only its imaging characteristics need to be
considered.  Like the lenses, the mirrors were only employed
before the optical beam was split or after the two paths had been
recombined such that wavefront distortions were common to both
paths.

After verifying the quality of our parallel-plate beamsplitters,
these optics were used in the evaluation of the other prisms.
Therefore our measurements were limited to the accuracy of the
beamsplitters.  It should be possible to remove this offset by
careful characterization of the beamsplitters used and by making
two sets of measurements with the beamsplitters flipped over
between them. But given the $\lambda/10$ surface quality of the
beamsplitters, it is possible that the deflection angles for two
different 1 cm sized spots on a beamsplitter will differ at the
microradian level even if the optic has no overall deflection
error. Lowering this systematic error, therefore, would require
either beamsplitters with better surface flatness or calibration
at the precise locations at which beams enter and leave the
beamsplitters.

\section{Conclusions}

In conclusion, we have demonstrated several relatively simple and
inexpensive techniques to characterize the deflection angles of
parallel-plate beamsplitters, penta prisms, right angle prisms,
and corner cubes.  We have achieved accuracies at the level of 2
$\mu$rad (0.4 arcseconds), approaching what is possible in
high-end commercial devices.  Better results are likely to be
possible by calibrating and removing effects due to imperfect
beamsplitters and by using a higher quality detector.

We acknowledge the contributions of Rebecca Merrill and Elizabeth
Cummings.  This work was supported by the Research Corporation and
the National Science Foundation.

\end{document}